\documentclass[11pt,english,twoside]{article}

\usepackage[T1]{fontenc}
\usepackage[latin1]{inputenc}
\usepackage[english]{babel}
\usepackage{lmodern}

\usepackage{amsmath,amssymb}

\usepackage{float}
\usepackage{graphicx}
\usepackage[dvips]{epsfig}
\usepackage{psfrag}
\usepackage{multicol}

\newcommand{\beq}{\begin{equation}}
\newcommand{\eeq}{\end{equation}}
\newcommand{\bea}{\begin{eqnarray}}
\newcommand{\eea}{\end{eqnarray}}
\newcommand{\nn}{\nonumber}
\newcommand{\w}{\wedge}
\newcommand{\sla}{\slash\!\!\!}

\begin{document}

\begin{flushright}

\end{flushright}

\begin{center}

{\Large \bf New supersymmetric flux vacua of type II string theory}

\vspace{.05in}

{\Large \bf and Generalized Complex Geometry}

\vspace{.2in}

{\Large David Andriot}

\vspace{.1in}

{LPTHE, CNRS, UPMC Univ Paris 06\\
Bo\^ite 126, 4 Place Jussieu\\
F-75252 Paris cedex 05, France}

\vspace{.1in}

{E-mail: \textit{andriot@lpthe.jussieu.fr}}

\vspace{.4in}

{\bf Abstract}

\end{center}

We study Minkowski supersymmetric flux vacua of type II string theory. Based on the work by M. Gra\~{n}a, R. Minasian, M. Petrini and A. Tomasiello, we briefly explain how to reformulate things in terms of Generalized Complex Geometry, which appears to be a natural framework for these compactifications. In particular, it provides a mathematical characterization of the internal manifold, and one is then able to find new solutions, which cannot be constructed as usual via T-dualities from a warped $T^6$ solution. Furthermore, we discuss how, thanks to a specific change of variables, one can ease the resolution of the orientifold projection constraints pointed out by P. Koerber and D. Tsimpis. One is then able to find new solutions with intermediate $SU(2)$ structure.

\vspace{.2in}

\textit{This paper contributes to the proceedings of the $4^{\textrm{th}}$ EU RTN Workshop, which took place during 09/2008 in Varna, Bulgaria. It is mainly based on references \cite{A} and \cite{GMPT}. Due to size constraints, several details and some citations are missing. For completeness, please have a look at \cite{A}.}

\vspace{.2in}

\section{Introduction}

To relate ten-dimensional string theory to real world four-dimensional low energy physics, a widely studied possibility is compactification. It consists at first in splitting space-time in a four-dimensional non-compact part (Minkowski $\mathbb{R}^{3,1}$ for us, but it can also be $dS_4$ or $AdS_4$), and a six-dimensional compact internal manifold $\mathcal{M}$. The properties of $\mathcal{M}$ influence in a non-trivial way the four-dimensional low energy physics obtained after dimensional reduction. So phenomenological criterion usually help to choose this manifold properly.

More precisely, one takes the low-energy effective theory of string theory, which is here ten-dimensional supergravity (SUGRA), and compactify it to get some four-dimensional supersymmetric (SUSY) quantum field theory. Trying to preserve some SUSY, actually the minimal amount, is a phenomenological requirement. This criteria leads to constraints on the possible $\mathcal{M}$. In the basic scheme, it turns out that $\mathcal{M}$ has to be a Calabi-Yau (CY) manifold. Choosing one CY, one can then compactify and try to compute the low-energy four-dimensional effective theory. Most of the time, one encounters the so-called moduli problem: several massless scalar fields, the moduli, are present in this effective theory. Some if not all should then be observed after SUSY breaking, but it is not the case. In recent years, a way to solve this problem has been developed: flux compactifications. The idea is to allow in the background for some (RR or NSNS) fluxes. These fluxes generate in the low energy effective action (LEEA) a potential for some of the moduli, giving them a mass! And flux compactifications have other nice virtues: they can generate hierarchies, they provide new possibilities for SUSY breaking, etc.

But adding fluxes can not be done for free: they change the SUSY conditions, so generically, $\mathcal{M}$ is no longer a CY. This can be simply understood: the fluxes and possible associated sources induce a backreaction on the manifold, so the latter cannot be flat anymore. The only difference is sometimes a warp factor: then we talk of a warped or conformal CY. The usual phenomenological approach is then to neglect, in a controlled way, this factor and with it the backreaction. But there are other situations where the backreaction is encoded in non-trivial fibrations, and then the manifold is much more different from a CY. So is it possible to determine in general on what $\mathcal{M}$ to compactify to preserve SUSY? For type II SUGRA, a mathematical characterization of $\mathcal{M}$ has been given in terms of Generalized Complex Geometry (GCG) \cite{H, G}: $\mathcal{M}$ preserving at least $\mathcal{N}=1$ has to be what is defined in GCG as a Generalized Calabi-Yau (GCY) \cite{GMPT04, GMPT05}.

Can we give explicit examples of backgrounds with fluxes? A first famous example consists in a warped $T^6$ with a three-dimensional orientifold (an O3-plane) and fluxes. Starting from this solution and performing T-dualities, one can get new flux vacua on non-CY manifolds \cite{KSTT}. There is another procedure, which uses the fact that $\mathcal{M}$ should be a GCY. Choosing such manifolds, one can perform a direct search for solutions on them. Doing so, one can find some vacua T-dual to warped $T^6$ solutions, but also some which are not. The latter are really ``new'' Minkowski SUSY vacua of type II SUGRA. A priori, they would not have been found without using GCG. It gives first indications that GCG appears as a good mathematical set-up to study type II SUGRA compactifications. Using these tools, we hope to get a better global understanding of the different possible vacua, and furthermore to get new possibilities for low-energy effective theories.

From now on we will focus on this second procedure to find vacua. We first come back to the different SUGRA vacua with particular emphasis on the SUSY conditions and the formalism. Then, we discuss the solutions found, mention difficulties about the O-plane projection constraints and their resolution, and finally give a non-trivial example of a ``new'' solution. We conclude with some opening remarks.

\section{Supergravity vacua and formalism}

\subsection{Supergravity vacuum}

Type II SUGRA has $\mathcal{N}_{10D}=2$. Its bosonic spectrum contains a metric $g$, a dilaton $\phi$, an NSNS flux $H$, and RR fluxes $F_p$. Its fermionic spectrum contains a doublet of gravitinos $\psi_{\mu \ 1,2}$ and a doublet of dilatinos
$\lambda_{1,2}$. We take the following ten-dimensional metric ansatz:
\beq
ds_{(10)}^2=e^{2A(y)}\ \eta_{\mu\nu}dx^\mu dx^\nu +g_{\mu\nu}(y) dy^\mu dy^\nu \ ,
\eeq
where $e^{2A}$ is the warp factor, $\eta_{\mu\nu}$ the Minkowski metric and $g_{\mu\nu}(y)$ the internal metric.

To find a Minkowski SUSY flux vacuum of type II SUGRA, one should first solve the bosonic equations of motion (e.o.m.). Once you allow for fluxes, one should furthermore verify their Bianchi identities (BI). Moreover, requiring that the vacuum should preserve some SUSY leads to some SUSY conditions that we are going to discuss. And finally, going to Minkowski leads to some other constraints, namely the tadpole cancellation, that we will come back to. One can show in our case that when both the SUSY conditions and the BI are verified, they imply together that the e.o.m. are automatically satisfied \cite{KT}. So our first concern will be about the SUSY conditions. Note that the same process is used when looking for SUGRA p-brane solutions.

The SUSY conditions consist in the annihilation of the SUSY variation of fermions in the vacuum\footnote{See \cite{A} for details.}:
\bea
\label{SUSY}
0&=&\delta \psi_\mu = D_\mu \epsilon + \frac{1}{4} H_\mu \mathcal{P} \epsilon + \frac{1}{16} e^{\phi} \sum_n \sla \! F_{2n} \gamma_{\mu} \mathcal{P}_n \epsilon \ , \nn\\
0&=&\delta \lambda = \left(\sla{\partial} \phi + \frac{1}{2} \sla \! H \mathcal{P}\right) \epsilon 
+ \frac{1}{8} e^{\phi} \sum_n (-1)^{2n} (5-2n)\ \sla \! F_{2n} \mathcal{P}_n  \epsilon \ ,
\eea
where the two Majorana-Weyl SUSY parameters of type II SUGRA fit in the doublet $\epsilon= (\epsilon^1,\epsilon^2)$. Note that the fluxes $H$ and $F_p$ are present in these SUSY conditions. If one puts them to zero, the first equation indicates the need for a covariantly constant spinor. This leads, after dimensional reduction, to the CY condition. With non-trivial fluxes, this condition is not obtained, so $\mathcal{M}$ is a priori not a CY.

The SUSY parameters can be decomposed as a product accordingly to the space-time splitting. In order to get $\mathcal{N}_{4D}=1$, we take the following generic decomposition
\bea
\label{decompo}
\epsilon^1 &=& \zeta \otimes \eta^1 + c.c. \ , \nn\\
\epsilon^2 &=& \zeta \otimes \eta^2 + c.c. \ ,
\eea
where $(\eta^1,\eta^2)$ is an internal pair of globally defined spinors\footnote{The chirality of the various spinors depend on the theory (IIA/B). We can choose to have $\eta^i$ of positive chirality and note it $\eta^i_+$. The opposite chirality is defined with the complex conjugation: $\eta^i_-=(\eta^i_+)^*$.}. For a consistent reduction to $\mathcal{N}_{4D}=1$, we need the existence of a pair of globally defined (nowhere vanishing) internal spinors $(\eta^1,\eta^2)$.

\subsection{Formalism: from internal spinors to Generalized Complex Geometry}

\begin{figure}[H]
\begin{center}
\begin{tabular}{cc|ccc}
\psfrag{eta1}[][b]{$\eta_+^1$}
\psfrag{eta2}[l][]{$\eta_+^2$}
\includegraphics[height=0.81cm]{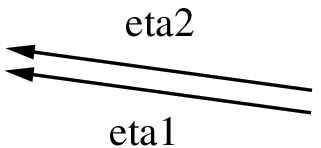}& & 

\psfrag{eta1}[][]{$\eta_+^1$}
\psfrag{eta2}[][]{$\eta_+^2$}
\psfrag{Phi}[][]{$\phi$}
\includegraphics[height=1.87cm]{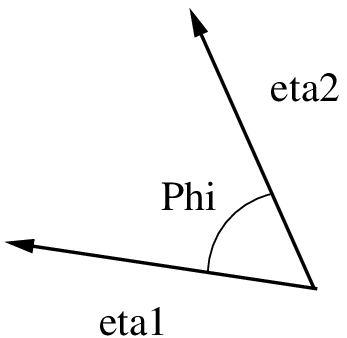}& &

\psfrag{eta1}[l][]{$\eta_+^1$}
\psfrag{eta2}[][]{$\eta_+^2$}
\includegraphics[height=2cm]{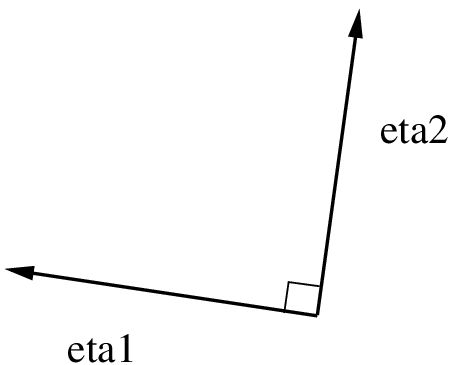}\\
 & & & & \\
$SU(3)$ structure & & Intermediate $SU(2)$ structure & & Static $SU(2)$ structure
\end{tabular}\caption{The different structures}\label{figstruc}
\end{center}
\end{figure}

The two spinors $(\eta^1_+,\eta^2_+)$ can be in different situations: either they are parallel, and then one talks of an $SU(3)$ structure, or they are not and then they define an $SU(2)$ structure. For the latter there are two subcases: either so-called static $SU(2)$ (``orthogonal'' spinors, where the ``orthogonality'' is defined with gamma matrices) or intermediate $SU(2)$ (a generic angle $\phi \neq \frac{\pi}{2}$). We sum-up the situations on the schematic pictures of figure \ref{figstruc}.
\vspace{-0.2in}
These $SU(3)$ or $SU(2)$ structures correspond to the structure group of the tangent bundle over $\mathcal{M}$, noted $T\mathcal{M}$. The number of independent globally defined spinors on $\mathcal{M}$ is related to the reduction of the structure group. For a CY threefold, we usually consider only one spinor which gives an $SU(3)$ structure.

Thanks to the Clifford map, we can construct globally defined differential forms, which differ according to the structure group/the spinors. For an $SU(3)$ structure, we can define the K\"ahler form $J$ (real two-form) and the holomorphic three-form $\Omega_3$, which appear usually for a CY. For an $SU(2)$ structure, we can define an holomorphic one-form $z$, a real-two form $j$ and an holomorphic two-form $\Omega_2$. We give their coefficients\footnote{See \cite{A} for details.}:
\vspace{-0.2in}
\begin{center}
\begin{tabular}{c|cc}
 & & $z_\mu = \eta_-^{\dagger} \gamma_\mu \chi_+ \qquad \qquad \qquad \qquad \qquad $ \\
$\qquad J_{\mu\nu} = - i \eta_+^{\dag}\gamma_{\mu\nu}\eta_+ \qquad $ & $\qquad \qquad \qquad \quad $ & $j_{\mu\nu} = - i \eta_+^{\dag}\gamma_{\mu\nu}\eta_+ +i \chi_+^{\dag}\gamma_{\mu\nu}\chi_+ \qquad \qquad \qquad \qquad \qquad $ \\

$\qquad \Omega_{\mu\nu\rho} = - i \eta_-^{\dag}\gamma_{\mu\nu\rho}\eta_+ \qquad $ & & $\Omega_{\mu\nu} = \eta_-^{\dagger} \gamma_{\mu\nu} \chi_- \qquad \qquad \qquad \qquad \qquad $\\

 & & \\

$SU(3)$ structure & & $SU(2)$ structure $\qquad \qquad \qquad \qquad \qquad $
\end{tabular}
\end{center}
where, in the following, $\eta_+$ is proportional to $\eta^1_+$, and $\chi_+$ is orthogonal to it. From now on, we can use these forms instead of the spinors\footnote{These forms have to satisfy some extra ``structure conditions'', to ensure the structure group is correctly reduced.}. In particular, the SUSY conditions can be rewritten in terms of the forms.

Let us go further with this logic. We introduce the bi-spinors:
\beq
\label{bispin}
\Phi_+ = \eta^{1}_+ \otimes \eta^{2\dag}_+\ , \quad \Phi_- = \eta^{1}_+ \otimes \eta^{2\dag}_- \ .
\eeq
Thanks to the Fierz identity, these are nothing but polyforms, i.e. sums of forms of different degree. They can be expressed in terms of the previously defined forms:
\begin{center}
\begin{tabular}{cc|ccc}
$\Phi_+/N_+ = e^{-iJ} $ & &  $ e^{\frac{1}{2} z\w \overline{z}} (c_{\phi}e^{-ij}-is_{\phi}\Omega_2)$ & & $-i \Omega_2 \w e^{\frac{1}{2} z \w \bar{z}} $ \\

$\Phi_-/N_- = i \Omega_3 $ & & $z\w (s_{\phi}e^{-ij}+i c_{\phi}\Omega_2)$ & & $z \w e^{-i j} $\\
 & & & & \\
$SU(3)$ structure & & Intermediate $SU(2)$ structure & & Static $SU(2)$ structure
\end{tabular}
\end{center}
where $N_{\pm}$ are normalisation factors and $c_{\phi}=\cos(\phi)$, $s_{\phi}=\sin(\phi)$. Once again we can use these polyforms instead of the spinors\footnote{As for the forms, the polyforms $\Phi_{\pm}$ have to satisfy some further constraints, the ``compatibility conditions'', to ensure the structure group is properly reduced.}, in particular for a rewriting of the SUSY equations.

These bi-spinors are of particular interest. We mentioned that each spinor $\eta^i_+$ defines an $SU(3)$ structure on the tangent bundle $T\mathcal{M}$. Then, one can show that the bi-spinors (\ref{bispin}) define an $SU(3)\times SU(3)$ structure on the bundle $T\mathcal{M}\oplus T^*\!\mathcal{M}$ where $T^*\!\mathcal{M}$ is the cotangent bundle. That is where Generalized Complex Geometry enters the game: $T\mathcal{M}\oplus T^*\!\mathcal{M}$ plays a crucial role in GCG. Furthermore, the bi-spinors $\Phi_{\pm}$ can be shown to be $\textrm{Cliff}(6,6)$ pure spinors on $T\mathcal{M}\oplus T^*\!\mathcal{M}$, and are related to generalised complex structures, important objects in GCG. So using these bi-spinors/polyforms/pure spinors $\Phi_{\pm}$ in this context is interesting because we can give interpretations in terms of GCG objects.

Using the decomposition (\ref{decompo}) and focusing on the internal part, the SUSY conditions (\ref{SUSY}) can be rewritten in terms of the pure spinors (\ref{bispin}) as\footnote{See \cite{A} for details.}
\bea
(d-H\w)(e^{2A-\phi}\ \Phi_1 )&=&0 \label{1eq}\\
(d-H\w)(e^{A-\phi}\ \textrm{Re}(\Phi_2 ))&=&0 \\
(d-H\w)(e^{3A-\phi}\ \textrm{Im}(\Phi_2 ))&=&\frac{e^{4A}}{8}*\lambda(\sum_p F_p ) \\
\textrm{with}\ \Phi_1=\Phi_\pm\ , \ \Phi_2=\Phi_\mp\ ,&&\textrm{IIA/B} \ .\nn
\eea
Note that the fluxes $H$ and $F_p$ appear in these equations in a rather simple way. Up to a factor, the pure spinor $\Phi_1$ is (twisted) closed in equation (\ref{1eq}), with respect to the exterior derivative $d$ ($-H\w$). This condition has an interpretation in GCG terms: a manifold carrying such a structure is called a (twisted) Generalized Calabi-Yau (GCY). Thanks to GCG, we finally get a mathematical characterization of the $\mathcal{M}$ preserving SUSY. From now on, looking for a solution, we know we have to take $\mathcal{M}$ to be a GCY.

\section{Solutions}

\subsection{Results}

We are now able to perform a direct search for Minkowski SUSY flux vacua instead of constructing them by T-duality. We just have to solve from scratch the equations discussed above (SUSY, BI...), on a given GCY. Let us give some results of this procedure.
\begin{itemize}
\item A systematic search for $SU(3)$ and static $SU(2)$ structure solutions was performed in \cite{GMPT}, among the 34 existing nilmanifolds and 13 solvmanifolds. These two kinds of manifolds are also known as twisted tori: they consist in fibrations of tori over other tori. Moreover, they are group manifolds. Importantly, the nilmanifolds have been proved to be GCY, so they are good candidates for the vacua we are interested in. From this systematic search, very few solutions were found. Among them, some were T-dual to a warped $T^6$ solution. Fortunately, some others were not: so truly new vacua were found. One was found on a nilmanifold $n\ 3.14$ and others on a solvmanifold $s\ 2.5$.

\item In \cite{A}, some intermediate $SU(2)$ structure solutions were found on those same two manifolds $n\ 3.14$ and $s\ 2.5$. Moreover, considering the angle $\phi$ of these solutions and taking its limit to the two extreme structures $SU(3)$ and static $SU(2)$, one could recover, at least partially, the solutions found previously in \cite{GMPT}. On the contrary to solutions found in \cite{KT}, these intermediate $SU(2)$ structure solutions are not T-dual to a warped $T^6$ solution, so they consist in truly new vacua.
\end{itemize}

Before giving an example of these solutions, let us come back to the tadpole cancellation and a difficulty appearing for intermediate $SU(2)$ structure solutions: the orientifold projection constraints.

\subsection{The orientifold projection}

In flux compactifications to Minkowski, there is a well known no-go theorem, also known as the tadpole cancellation: one has to cancel the contributions of the fluxes and the source charges to the trace of the energy-momentum tensor \cite{GKP}. To do so, one needs negatively charged sources: the O-planes (orientifolds). But one cannot bring for free an O-plane into the game. An O-plane induces a projection on states, so a solution has to satisfy some additional constraints. In particular, it has to transform in a precise manner under a target space involution $\sigma$. For instance, the SUSY internal parameters or the structure forms have to transform this way \cite{KT}:
\begin{center}
\begin{tabular}{rc|cl}
$O5\ :\ \sigma(\eta_{\pm}^1)=\eta_{\pm}^2$ & $\quad \qquad $ & $\qquad \quad $ & \\
$\sigma(\eta_{\pm}^2)=\eta_{\pm}^1$ & & & $\qquad \ \ \sigma(j) = c_{2\phi}\ j+ s_{2\phi}\ \textrm{Re}(\Omega_2)$ \\
$O6\ :\ \sigma(\eta_{\pm}^1)=\eta_{\mp}^2$ & & & $\sigma(\textrm{Re}(\Omega_2)) = s_{2\phi}\ j - c_{2\phi}\ \textrm{Re}(\Omega_2)$\\
$\sigma(\eta_{\pm}^2)=\eta_{\mp}^1$ & & & $\sigma(\textrm{Im}(\Omega_2)) = - \textrm{Im}(\Omega_2)$
\end{tabular}
\end{center}
where $c_{2\phi}=\cos(2\phi)$, $s_{2\phi}=\sin(2\phi)$. For $SU(3)$ or static $SU(2)$ structures ($\phi=0, \ \frac{\pi}{2}$), these conditions are rather simple to verify. But for a generic intermediate $SU(2)$ structure solution, they become non-trivial constraints.

\begin{figure}[H]
\begin{center}
\psfrag{Psi}[][]{$ $}
\psfrag{eta1}[][]{{\small $\eta_+^1$}}
\psfrag{eta2}[][]{{\small $\eta_+^2$}}
\psfrag{etaD}[t][]{{\small $\eta_{+D}$}}
\psfrag{etaDo+}[l][l]{{\small $\frac{z.\eta_{-D}}{2}$}}
\includegraphics[height=2cm]{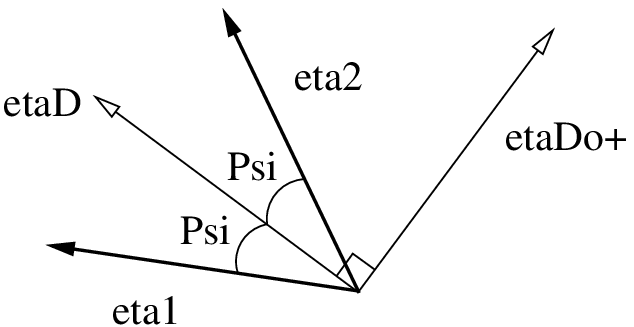}\caption{The dielectric spinor}\label{diel}
\end{center}
\end{figure}

There is a way-out thanks to a specific change of variables: one can choose the structure defining spinors to be the dielectric spinor $\eta_{+D}$ and the one orthogonal to it. $\eta_{+D}$ is proportional to the bi-sector spinor of the SUSY parameters, as pictured on figure \ref{diel}. The projection conditions on this spinor and the associated structure forms (IIA/B) are much simpler \cite{A}:
\begin{multicols}{3}
\columnbreak
\bea
O5&:& \sigma(\eta_{\pm D})=\eta_{\pm D} \nn\\
O6&:& \sigma(\eta_{\pm D})=\eta_{\mp D} \nn \\
&& \nn
\eea
\columnbreak
\bea
&& \nn \\
\sigma(j_D) &=& \mp j_D \nn\\
\sigma(\textrm{Re}(\omega_D)) &=& \pm \textrm{Re}(\omega_D) \nn\\
\sigma(\textrm{Im}(\omega_D)) &=& -\textrm{Im}(\omega_D) \nn\\
&& \nn
\eea
\end{multicols}
\vspace{-0.3in}
The SUSY equations are simplified as well. So these are the proper variables to use for intermediate $SU(2)$ structure solutions.

To sum-up, the orientifold projection has to be taken into account. It can be a non-trivial requirement, but then it can be simplified by a specific change of variables.

\subsection{Example of solution}

Let us present briefly an intermediate $SU(2)$ structure solution of type IIB found on the solvmanifold $s\ 2.5$ \cite{A}. As a group manifold, it is defined by the algebra $(25,-15,\alpha 45,-\alpha 35, 0, 0)$, $\alpha = \pm1$. It can be understood geometrically as some multiple fibration of circles. The solution has an O5 and a D5 along directions 13 and 24. It cannot be T-dualised to a warped $T^6$ solution. We first give the solution in terms of the $SU(2)$ structure forms, developed on a basis of real one-forms $e^{i=1..6}$, and expressed in terms of the complex structure moduli $\tau_{1,3,4,5,6}$ and the angle $\phi$:
\begin{center}
\begin{tabular}{l|l}
$z=\tau_5 e^5+i \tau_6 e^6$ & $\quad z^1=\tau_1 e^1- i \tau_3 e^3 +\tau_4 e^4$\\
$\Omega_2= z^1\w z^2$ & $\quad z^2=e^2 - \frac{\alpha \tau_3^2 s_{\phi}^2}{\tau_1\tau_4}e^3-i\frac{\alpha \tau_3}{\tau_1}e^4$ \\
$j=\frac{i}{2} ( t_1 z^1\w \overline{z}^1 + t_2 z^2\w \overline{z}^2+ b z^1\w \overline{z}^2 - \overline{b} \overline{z}^1 \w z^2 ) \quad$ & $\quad t_1=\frac{\alpha \tau_3 s_{\phi}}{\tau_1\tau_4 c_{\phi}}=\frac{1}{t_2 c_{\phi}^2},\ b=i\frac{s_{\phi}}{c_{\phi}}$
\end{tabular}
\end{center}
with holomorphic one-forms $z^i$ and K\"ahler moduli $t^i$ and $b$. We can relate these forms to the vacuum spectrum with $e^{\phi}=g_s=\textrm{cst}$, $H=0$, $F_1=0$, $F_5=0$, and
\beq
g=\textrm{diag} \left (\frac{\alpha \tau_3 \tau_1 s_{\phi}}{\tau_4 c_{\phi}} ,\ \frac{\alpha \tau_4 \tau_1}{\tau_3 s_{\phi} c_{\phi}},\ \frac{\alpha \tau_3^3 s_{\phi}c_{\phi}}{\tau_1 \tau_4},\ \frac{\alpha \tau_3 \tau_4 c_{\phi}}{\tau_1 s_{\phi}},\ \tau_5^2,\ \tau_6^2 \right )\ , \ F_3=-\frac{e^{-i\theta} (\tau_4^2-\tau_3^2 s_{\phi}^2)\ |\tau_6|}{g_s \tau_4|\tau_5| s_{\phi}} (e^2 \w e^3 \w e^6+\alpha\ e^1 \w e^4 \w e^6) \ .\nn
\eeq
Taking the limit of $\phi$ to the $SU(3)$ or static $SU(2)$ structure cases is possible, but is more intricate than it seems, because some of the moduli vary as well. Nevertheless, one recovers at least partially the solutions of \cite{GMPT}.

\section{Conclusions}

We discussed briefly how GCG appears as a natural framework for SUSY compactifications of type II SUGRA. In particular, it provides a mathematical characterization of $\mathcal{M}$ as a GCY. One is then able to perform a direct search for vacua. Some of the vacua found are T-duals to warped $T^6$ solutions and others are not. The latter are truly new vacua, that one would a priori not have found without GCG. Some of these new vacua carry $SU(3)$ or static $SU(2)$ structures \cite{GMPT}. Thanks to an additional reformulation of the orientifold projection constraints, new intermediate $SU(2)$ structure solutions were found as well \cite{A}. Taking the limit on these new solutions to $SU(3)$ and static $SU(2)$ structure cases, one recovers at least partially the solutions found in \cite{GMPT}.

GCG points out another possible kind of solutions: the dynamical $SU(3)\times SU(3)$ solutions. These are coordinate dependant solutions, where in particular the coordinate dependence can be in the angle $\phi$. This angle could evolve from point to point and the structure could become $SU(3)$ or static $SU(2)$ at some points. No such solution has been found in the compact case so far. As these solutions are ``at most points'' of the intermediate $SU(2)$ structure kind, one could hope that the techniques developed in \cite{A} could be useful to find some.

GCG is a natural framework for the study of non-geometry as well. The frontier between geometry and non-geometry appears clearly in this context \cite{GMPW}. Once again, GCG seems to be a natural framework for flux compactifications, the fluxes being geometric or not.

A motivation for this work has been to understand better the different possible Minkowski SUSY flux vacua of type II SUGRA. A natural phenomenological question that follows is of course that of the LEEA on these vacua. This is actually a tough point, which is still under study. See for instance \cite{GLW}.\\

\textbf{Acknowledgements}\\

I would like to thank my PhD supervisor Michela Petrini for her help and advices on this work. It is a pleasure to thank the organisers of the $4^{\textrm{th}}$ EU RTN Workshop for giving me the opportunity to present this work, and for the nice time we had there. I acknowledge support by ANR grant BLAN05-0079-01.

\end{document}